\documentclass[apj]{emulateapj} 
\usepackage{color}
\usepackage[english]{babel}
\usepackage{amsmath,amsfonts,amssymb}
\usepackage{xspace,multirow}
\usepackage{lineno}
\newcommand{\dd}{{\rm d}}

\newcommand{\asr}{Adv. Space Res.}

\newcommand{\gu}{\ensuremath{\gamma_{1}}}
\newcommand{\gd}{\ensuremath{\gamma_{2}}}

\newcommand{\RH}{Rankine-Hugoniot\xspace}

\newcommand{\mr}[1]{\multirow{2}{*}{#1}}

\shorttitle{Multi-polytropy and entropy}
\shortauthors{Scherer et al.}

\begin{document}

\title{Generalized multi-polytropic Rankine-Hugoniot relations and the
  entropy condition}

\author{Klaus Scherer\altaffilmark{1}  and  Horst Fichtner\altaffilmark{1}}
\affil{Institut f\"ur Theoretische Physik IV: Weltraum- und
  Astrophysik, Ruhr-Universit\"at Bochum, Germany}
\email{kls@tp4.rub.de, hf@tp4.rub.de}
\altaffiltext{1}{Research Department, Plasmas with Complex Interactions,
  Ruhr-Universit\"at Bochum, Germany}
\author{Hans J\"org Fahr} 
\affil{Argelander Institut f\"{u}r Astronomie, Universit\"{a}t Bonn, Germany}
\email{hfahr@astro.uni-bonn.de}
\author{Christian R\"oken} 
\affil{Universit\"at Regensburg, Fakult\"at f\"ur Mathematik,
  Regensburg, Germany}
\email{christian.roeken@mathematik.uni-regensburg.de}
\author{Jens Kleimann} 
\affil{Institut f\"ur Theoretische Physik IV: Weltraum- und
  Astrophysik, Ruhr-Universit\"at Bochum, Germany}
  \email{jk@tp4.rub.de}

\begin{abstract}
  The study aims at a derivation of generalized \RH relations,
  especially that for the entropy, for the case of different
  upstream/downstream polytropic indices and their implications.  We
  discuss the solar/stellar wind interaction with the interstellar
  medium for different polytropic indices and concentrate on the case
  when the polytropic index changes across hydrodynamical shocks. We
  use first a numerical mono-fluid approach with constant polytropic
  index in the entire integration region to show the influence of the
  polytropic index on the thickness of the helio-/astrosheath and on
  the compression ratio. Second, the \RH relations for a polytropic
  index changing across a shock are derived analytically, particularly
  including a new form of the entropy condition. In application to
  the/an helio-/astrosphere, we find that the size of the
  helio-/astrosheath as function of the polytropic index decreases in
  a mono-fluid model for indices less than $\gamma=5/3$ and increases
  for higher ones and vice versa for the compression
  ratio. Furthermore, we demonstrate that changing polytropic indices
  across a shock are physically allowed only for sufficiently high
  Mach numbers and that in the hypersonic limit the compression ratio
  depends only on the downstream polytropic index, while the ratios of
  the temperature and pressure as well as the entropy difference
  depend on both, the upstream and downstream polytropic indices.
\end{abstract}


\section{Introduction}

The concept of a polytropic equation of state that relates the thermal
pressure $P$ to the mass density $\rho$ via
\begin{eqnarray}\label{EOS}
  P \rho^{-\tilde{n}} = \rm{const}\ ,
\end{eqnarray}
where $\tilde{n}$ is the so-called polytropic index, has, since its
introduction by \citet{Zeuner-1887}, found innumerable applications in
physics and engineering, reaching from the general behavior of
astrophysical gases \citep{Horedt-2004} and the structure of
stars \citep{Emden-1907, Chandrasekhar-1939}, to technical and
laboratory gases \citep{Drake-2006, Moran-etal-2014}. The usefulness
of the polytropic description arises from both its simplicity and its
validity for a wide range of rather different physical processes and
scenarios.  Depending on the choice of polytropic index,
Eq.~(\ref{EOS}) can describe isobaric ($\tilde{n}=0$), isothermal
($\tilde{n}=1$), isentropic ($\tilde{n}=\gamma$ = ratio of specific
heats), or isochoric ($\tilde{n}\rightarrow \infty$) processes.  Often,
$\gamma$, rather than $\tilde{n}$, is referred to as the polytropic index, and
we adopt this convention here. The actual value of $\gamma$ for a
specific system depends on its degrees of freedom $f$: For the
non-relativistic case, one has $\gamma = (f+2)/f$, whereas for the
highly relativistic case \citep[see, e.g.,][]{Horedt-2004}, one has
$\gamma = (f+1)/f$.  The fact that $f$ is a positive integer implies
that the physically meaningful $\gamma$ values must satisfy
$1 < \gamma \le 3$ for the non-relativistic case, and
$1 < \gamma \le 2$ in the highly relativistic limit.

Particularly in astrophysical applications, a non-constant polytropic
index is often considered. A classical example is a variable
polytropic index in models of stellar structure
\citep{Eddington-1938}, which have been termed {\it multi-polytropic}
models \citep{Buitrago-Calco-Mozo-2010, Livadiotis-McComas-2013}. Also
space plasmas, like the solar wind, have been described by means of a
variable polytropic index
\citep[e.g.,][]{Fahr-etal-1977, Totten-etal-1996, Fahr-2002b,
  Fahr-2002a, Fahr-Rucinski-2002, Chashei-etal-2003,
  Roussev-etal-2003}.
The advantage of such a description is that the non-thermal heating,
which is required to counteract the adiabatic cooling of the expanding
wind plasma, can be simulated without specifying an explicit physical
heating process.  For the same reason, constant $\gamma$ values of ---
or slightly larger than --- unity are often used to model the solar
wind plasma in the inner heliosphere
\citep[e.g.,][]{Parker-1958, Keppens-Goedbloed-1999,
  Kleimann-etal-2009}.
It is worth noting that in these cases, $\gamma$ is chosen to
heuristically describe a large-scale energy balance, rather than a
microscopic material property of the plasma. For instance,
$\gamma=1.05$ would imply $f=40$ degrees of freedom, which is clearly
unrealistic for any known state of matter.  Moreover, such small
values of $\gamma$ yield vastly incorrect jump conditions at shocks.
For an in-depth discussion of this problem, as well as an appropriate
method to address it in the context of solar wind simulations, see
\citet{Pomoell-etal-2011} and \citet{Pomoell-Vainio-2012}.

Considering composite gases or plasmas, the different
constituents will, in general, be characterized by different polytropic
indices \citep[e.g.,][]{Wu-etal-2009}. This scenario was also referred to as
multi-polytropic by \citet{Fahr-Siewert-2015} in their study of multi-fluid
magnetohydrodynamic (MHD) shocks with emphasis on the solar wind termination
shock (TS). These authors, however, for simplicity used a mono-fluidal, rather
than a multi-fluid description of the plasma, implying a change of the
polytropic index across the TS.

Systems with varying polytropic indices upstream and downstream of a
shock have recently been studied for the inner heliosheath (IHS),
i.e., the region of subsonic solar wind plasma between the TS and the
heliopause (HP), by \citet{Izmodenov-etal-2014}. The authors mimicked
an effective heat conduction in the IHS by a reduced polytropic index
of $\gamma = 1.06$, thus obtaining a nearly isothermal plasma state.
While they did not consider a jump of the polytropic index across the
TS, they did so for the HP: At this tangential discontinuity, $\gamma$
jumps from 1.06 to $5/3 \approx 1.67$.

As a consequence of having different polytropic indices on either side
of a discontinuity, the \RH relations have to be generalized. Such
generalized \RH relations have been derived by
\citet{Nieuwenhuijzen-etal-1993} including ionization, dissociation,
radiation, and related phenomena such as excitation, rotation, and
vibration of molecules, and, e.g., by \citet{Drake-2006} and
\citet{Livadiotis-2015}.   However, these studies remain
  incomplete because they did not derive an explicit
expression for the resulting entropy change.

Here, we derive the complete \RH relations for a single-species
hydrodynamic fluid for which we assume that the polytropic index on
both sides of a shock differs: $\gamma_{1}$ in the unshocked
supersonic wind region and $\gamma_{2}$ in the subsonic IHS or the
astrosheath. To this end, we first recall results for the solar wind
TS in Section~\ref{sec:2}. In Section~\ref{sec:3}, we investigate the
case of globally constant polytropic indices in systems with
shocks. Generalized \RH relations arising from a change of the
polytropic index across a shock are extensively discussed in
Section~\ref{sec:4}, and their detailed derivation is provided in the
Appendix. Section~\ref{sec:5} analyzes the entropy change over the
shock and its consequences. This is of paramount importance since the
entropy condition always plays a crucial rule in deriving the allowed
shock transitions
\citep{Courant-Friedrichs-1948, Goedbloed-etal-2010}. We here derive
for the first time the entropy difference for the case of different
polytropic indices on either side of the shock. All results are
summarized in the concluding Section~\ref{sec:6}. Throughout the
paper, major results are applied to the heliosphere and to
astrospheres, respectively.
~\\

\section{The solar wind termination shock}
\label{sec:2}

\subsection{The need for variable polytropic indices across the shock}
\label{sec:22}

A motivation to consider variable polytropic indices across shocks
originates from the study of multispecies and multi-fluid plasmas.  In
a so-called magneto-adiabatic study, \citet{Fahr-Siewert-2015} found
that the pressures $P_{\rm pui, 1/2}$ of pickup ions (PUIs) upstream
and downstream of the shock, as derived in a multi-fluid approach, are
related to the corresponding solar proton pressures $P_{\rm p, 1/2}$
by
\begin{eqnarray}
  \label{F1}
  \frac{P_{\rm pui ,2}}{P_{\rm pui, 1}} = \frac{P_{\rm p ,2}}{P_{\rm p, 1}}
  = \frac{s}{3} \left( 2 A(\alpha, s) + \frac{s^2}{A^2(\alpha, s)} \right)
 \, ,
\end{eqnarray}
where $\alpha$ denotes the tilt angle between the upstream magnetic
field and the shock surface normal,
$s:= \rho_{{\rm pui}, 2}/\rho_{{\rm pui}, 1}$ is the compression ratio
(of downstream to upstream PUI mass densities), and
\mbox{$A(\alpha, s) := \sqrt{\cos^{2}\alpha + s^{2} \sin^{2}\alpha}$}
\citep{Fahr-Siewert-2013}. Throughout this paper, subscripts~1 and~2
denote upstream and downstream values, respectively.  In an
alternative approach to the problem of multi-fluid shocks,
\citet{Wu-etal-2009} obtained the pseudo-polytropic relation
\begin{eqnarray}
  \label{eq:Prat_wu}
  \frac{P_{\rm pui,2}}{P_{\rm pui,1}}
  = \left( \frac{ \rho_{\rm pui ,2}}{\rho_{\rm pui,1}} \right)^{\tilde{\gamma}}
  = s^{\tilde{\gamma}}
\end{eqnarray}
for the downstream and upstream PUI pressures, in which a PUI-specific
polytropic index $\tilde{\gamma} \ge 5/3$ is used to describe a particular
heating for PUIs at the shock passage in order to obtain better agreement
between the simulation results and the {\em Voyager~2} shock data shown by
\citet{Richardson-etal-2008}. 

Combining formulas (\ref{F1}) and (\ref{eq:Prat_wu}) yields \citep{Fahr-Chalov-2008}
\begin{eqnarray}
\label{eq:s}
  s^{\tilde{\gamma}} 
  = \frac{s}{3} \left( 2A(\alpha,s)+\frac{s^{2}}{A^{2}(\alpha,s)}\right) \, .
\end{eqnarray}
For a perpendicular shock, i.e., $\alpha =90^{\circ}$, the right hand
side of Eq.~(\ref{eq:s}) simplifies to $(s/3)(2s+1)$.  For a
compression ratio of, say, $s=3$, one obtains $A(90^{\circ},3)$ and,
hence,
\begin{eqnarray}
  \tilde{\gamma}(90^{\circ}, 3)=\frac{\ln (7)}{\ln (3)}
  \approx 1.77 > \frac{5}{3} \, .
\end{eqnarray}
In the case of a quasi-parallel shock with $\alpha =20^{\circ}$ and again
$s = 3$, one analogously obtains $\tilde{\gamma}(20^{\circ},
3)=2.1$. Moreover, 
a well-known example for fluid motion with only two degrees of freedom is
that of shallow water waves.

We remark that the introduction of a polytropic index $\tilde{\gamma}$ in the
latter theoretical approach invokes an unexplained ad~hoc process for PUIs,
since it treats the PUI protons with respect to their thermodynamic response
to shock compression substantially different than the normal solar wind
protons. It may be more reasonable to think that ``protons are protons'',
even if they are called ``PUIs''.

Another motivation to study various polytropic indices arises
from having systems with different degrees of freedom in different
directions, which occurs in, e.g., magnetized plasmas due to the different
behaviors of ions parallel and perpendicular to the magnetic field.
\citet{Siewert-Fahr-2007} considered the so-called CGL plasma
invariants at a quasi-perpendicular TS, and obtained for the
conversion from upstream to downstream pressure components parallel
and perpendicular to the magnetic field the relations
\begin{eqnarray}
  P_{\parallel ,2} &=& s \, P_{\parallel ,1} \\
  P_{\perp ,2} &=& s^{2} P_{\perp ,1} \, .
\end{eqnarray}
Interpreting these as a polytropic reaction of the plasma ions due to
direction-specific polytropic indices $\gamma_{\parallel}$ and
$\gamma_{\perp }$ leads to
\begin{eqnarray}
  P_{\parallel,2}/P_{\parallel,1}&=& s=(\rho_2/\rho_1)^{\gamma_{\parallel}} \\
  P_{\perp,2}/P_{\perp,1}       &=& s^{2}=(\rho_2/\rho_1)^{\gamma_{\perp}}
\end{eqnarray}
with polytropic indices $\gamma_{\parallel}=1$ and $\gamma_{\perp}=2$.
For quasi-parallel shocks, the corresponding indices are instead
$\gamma_{\parallel}=3$ and $\gamma_{\perp}=1$ \citep[see][]{Siewert-Fahr-2008}.
~\\

\subsection{The need for a generalized entropy jump
  formula}
\label{sec:20}

The entropy of an ideal gas, which is an extensive quantity, is often
expressed in terms of a fixed polytropic index across a shock, namely
by the quantity $\hat{S}_{i} := P_{i} \rho_{i}^{-\gamma_{i}}$
\citep[see, e.g.,][]{Goedbloed-etal-2010}. Along a streamline, this
yields for the mono-fluid approximation
\begin{eqnarray}
  \hat{S}_{1} &=&
  \frac{P_{\rm pui,1}+P_{\rm p,1}}{(\rho_{\rm pui,1}+\rho_{\rm
                  p,1})^{\gamma _{1}}}ß ,
  \\ \nonumber \\
  \hat{S}_{2} &=& \frac{P_{\rm pui,2}+P_{\rm p,2}}{(\rho_{\rm pui,2}+\rho_{\rm p,2})^{\gamma_2}} \ .
\end{eqnarray}
Obviously, $\hat{S}_{1}$ and $\hat{S}_{2}$ have, in general, different
dimensions. This shows that these quantities cannot be used
straightforwardly to determine the entropy change across a shock.
Employing the Sackur-Tetrode formula \citep{Sackur-1913} leads to a
similar problem since on either side of the shock a different
normalization constant for the entropy arises depending on the
polytropic index \citep[see, e.g.,][]{Oliveira-2013}.  Thus, neither
the quantities $\hat{S_{i}}$ nor the Sackur-Tetrode formula can be
applied.

One has to reconsider the concept of the entropy change across a
multi-polytropic shock from a more fundamental perspective. A thorough
discussion is given in Section~\ref{sec:5}.
~\\

\section{Globally constant polytropic indices}
\label{sec:3}

\begin{figure}[t!]
  \centering
  \includegraphics[width=0.95\columnwidth]{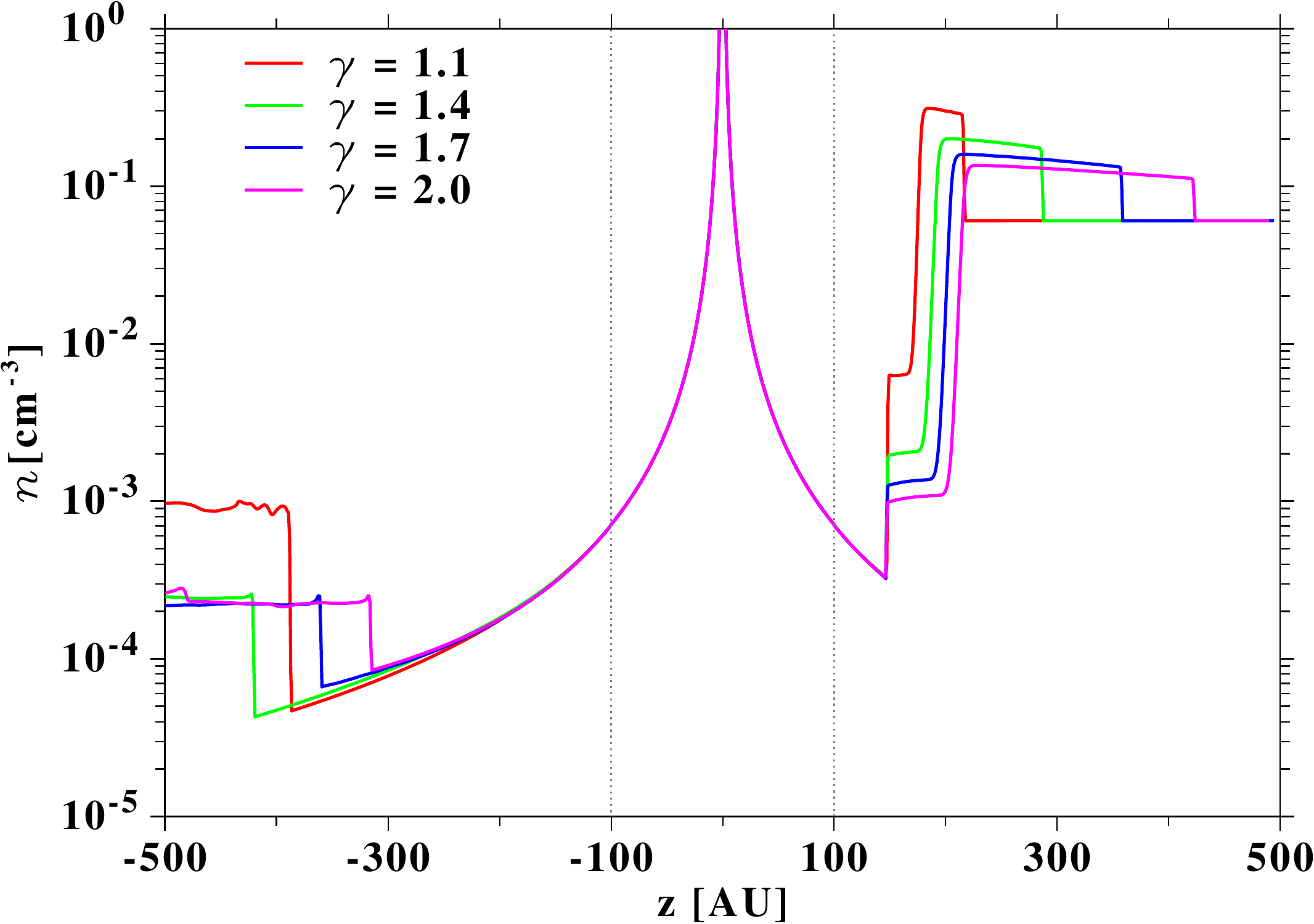}
  \caption{\label{fig:1} The number density along the ISM inflow axis
    in upwind \mbox{($z>0$)} and downwind \mbox{($z<0$)} direction in
    the steady-state configuration of the simulation. The two dotted
    vertical lines at $z=\pm 100$~AU mark the inner boundary of the
    computational domain, inside of which the density was estimated
    according to \mbox{$\rho\propto r^{-2}$}.  The first jump in
    upwind direction is the termination shock, the second one is the
    heliopause, and the third one is the bow shock.  In the downwind
    direction, there is only a termination shock.}
\end{figure}

Before considering the case of variable polytropic indices across a
shock, we first discuss the one of globally constant polytropic
indices in order to illustrate the general influence of polytropic
indices on shock structures. As stated above, this is done with the
examples of the heliosphere and of astrospheres, whose large-scale
structures are often described with hydrodynamic models
\citep[e.g.,][]{Pauls-Zank-1996, Fahr-etal-2000,
  Borrmann-Fichtner-2005, Mueller-etal-2008, Arthur-2007,
  Izmodenov-etal-2014, Scherer-etal-2015a}.
In Fig.~\ref{fig:1}, we show the number densities along the inflow
axis of the interstellar medium (ISM) for different polytropic indices
based on our simulations with the Cronos MHD code
\citep{Kissmann-etal-2008, Kleimann-etal-2009, Wiengarten-etal-2015,
  Scherer-etal-2015a, Scherer-etal-2016}.
We have modeled the different polytropic scenarios with the Cronos
hydrodynamic one-fluid module for the parameters used in the
\citet{Mueller-etal-2008} benchmark: The boundary values at 1~AU are
$v_{\rm sw}= 375$~km\,s$^{-1}$, $n_{\rm sw}=7$~cm$^{-3}$, and
$T_{\rm sw}=7.364 \times 10^{4}$~K, whereas those for the ISM are
$v_{\rm ism}=26.4$~km\,s$^{-1}$, $n_{\rm ism}=0.06$~cm$^{-3}$, and
$T_{\rm ism}=6.53 \times 10^{3}$~K. As usual, an axisymmetric
configuration forms which is characterized by the TS terminating the
supersonic solar/stellar wind, a bow shock (BS) on the upwind side,
where the interstellar flow (as seen in the rest frame of the
Sun/star) changes from supersonic to subsonic, and a tangential
discontinuity, the helio-/astropause (HAP) in between. The inner
helio-/astrosheath (IHAS) is the region between the TS and the HAP,
and the outer helio-/astrosheath (OHAS) that between the HAP and the
BS.

At the inner computational boundary of the model (located at a
helio-/astrocentric distance of $r=r_{\rm b} = 100$~AU), the upstream
temperature $T_{1}$, the thermal pressure $P_{1}$, and the Mach number
$M_{1}=v_1/c_1$ are extrapolated inwards from their respective
boundary values using the polytropic relation $P_1 \propto \rho_1^{\gamma}$
and the ideal gas law with
\begin{eqnarray}
  \label{eq:extrapol}
  P_{1} \propto r^{-2 \gamma} \, , \quad T_{1} \propto r^{-2(\gamma-1)} \, ,
  \quad M_{1} \propto r^{\gamma-1} \, .
\end{eqnarray}
Moreover, the Mach number at the TS distance can easily be estimated
according to formulas (\ref{eq:extrapol}).

\begin{table}
  \begin{tabular}{c|rrrr|l}
    $\gamma$ & $c_{\rm s}$ [km/s] & $M$ & $M$ & $s$ &  \\   
    & 100~AU & 100~AU &  150~AU &  \\
    \hline
    \mr{1.06} & \mr{19.3} & \mr{19.1} & \mr{20.1} & \mr{33.5}
    & Izmodenov et~al. \\ &&&&& (2014)\\
    1.10 & 16.2 & 23.2 & 24.1 & 20.6 & (Fig.~\ref{fig:1}) \\
    4/3  &  6.1 & 61.3 & 70.2 &  7.0 & electron gas \\
    \mr{7/5} & \mr{4.5} & \mr{82.7} & \mr{97.5} & \mr{5.9} & diatomic gases\\
    &&&&& (Fig.~\ref{fig:1}, $\gamma\approx 1.4$) \\
    \mr{5/3} & \mr{1.5} & \mr{254.2} & \mr{333.1} & \mr{4.0}
    & monoatomic gases \\
    &&&&& (Fig.~\ref{fig:1}, $\gamma \approx 1.7$) \\
    \mr{2} & \mr{0.4} & \mr{1075.5} & \mr{1613.3} & \mr{3.0} &
    shallow water waves \\
    &&&&& (Fig.~\ref{fig:1}) \\
  \end{tabular}
  \caption{ \label{tab:1} \rm \begin{flushleft}
      The first column shows the polytropic index, the second and third the
      sound speed and Mach number at 100~AU, and the fourth column the Mach
      number at the TS at 150~AU. In the fifth column, the corresponding
      compression ratio is given.
      Note again the unrealistically high values of $s$ in the first two rows.
    \end{flushleft}
  }
\end{table}

\begin{figure}[t!]
  \centering
  \includegraphics[width=0.95\columnwidth]{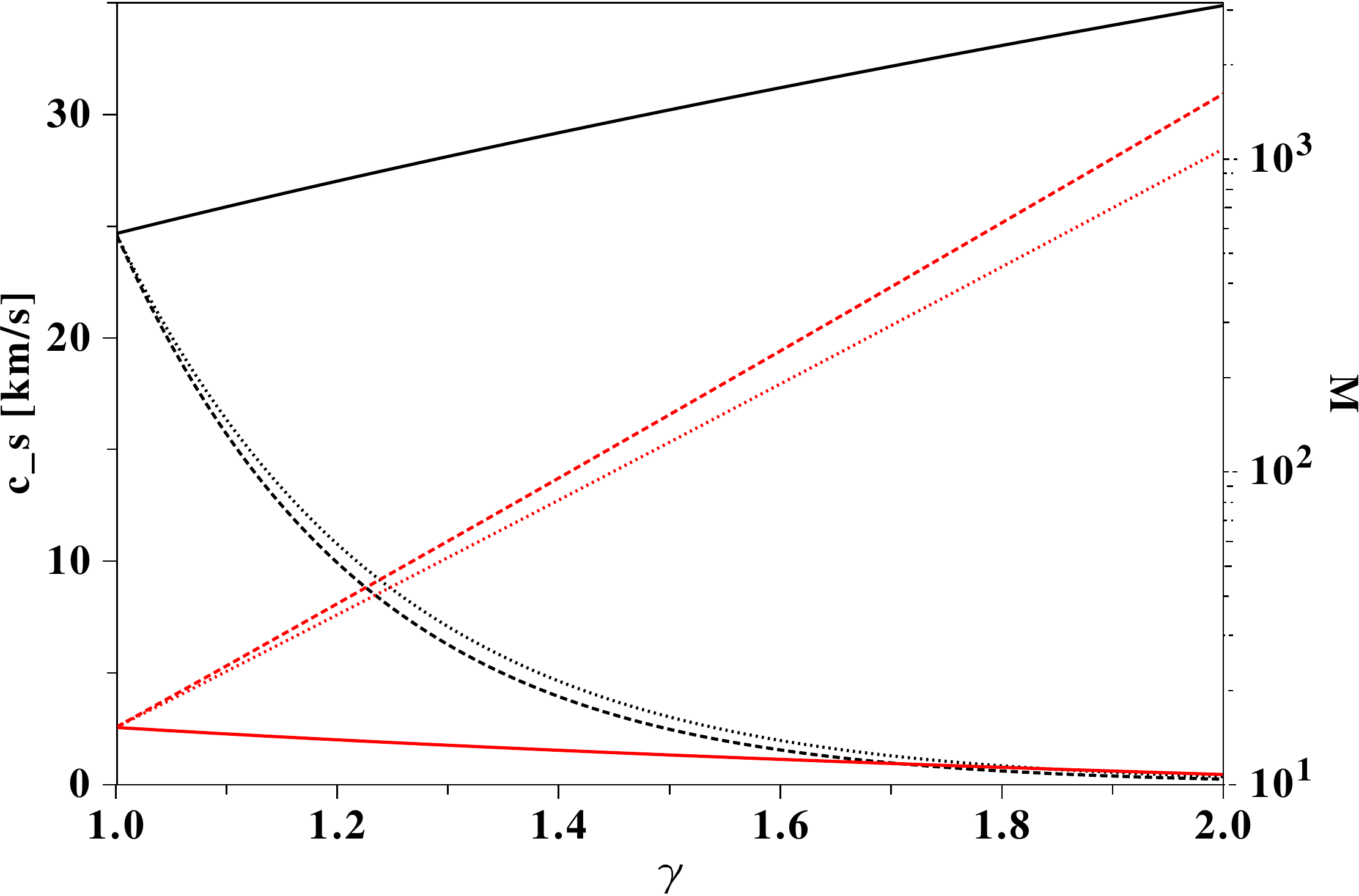}
  \caption{ \label{fig:0}
    The sound speeds (black) and Mach numbers (red) as functions of the
    polytropic index. The left y~axis shows the sound speed at 1~AU
    (solid line), 100~AU (dotted line), and at 150~AU (dashed line).
    The right y~axis shows the logarithm of the Mach number, where the
    solid, dotted, and dashed lines correspond to the above. For an
    increasing polytropic index, the sound speed at 1~AU increases, whereas
    at 100 and 150~AU it decreases. The Mach number shows the opposite
    behavior.
 }
\end{figure}

The sound speeds and Mach numbers for different polytropic indices are
shown in Fig.~\ref{fig:0} at 1~AU, 100~AU, and 150\,AU (which is the
TS distance in upwind direction). From the definition of the sound
speed
\begin{equation}
  c_{\rm s} := \sqrt{\frac{\gamma \, k T}{m_{\rm p}}} \, ,
\end{equation}
where $k$ is the Boltzmann constant, $m_{\rm p}$ the proton mass, and
$T$ given by Eq.~(\ref{eq:extrapol}), it follows that
\mbox{$c_{s}|_{1\,{\rm AU}} \propto \sqrt{\gamma}$}, while already for
slightly larger radii, and particularly for 100 and 150~AU, the
$\gamma$ dependence in $T$ dominates, leading to an exponential decay.
Consequently, the Mach numbers at 1\,AU, 100~AU, and 150~AU show the
corresponding inverse behaviors.  In Table~\ref{tab:1}, specific
values for sound speed, Mach number, and compression ratio are given
for different $\gamma$ values. This table shows that at the TS the
hypersonic approximation $M \gg 1$ (see Appendix) can be applied to
the \RH relations. 

Caution has to be taken when choosing the polytropic indices for
specific systems: for small $\gamma$'s (i.e., close to unity) the
kinetic energy is increasingly transferred into the downstream ram
pressure rather than in thermal pressure of the plasma. Moreover, the
polytropic index has to be expressed by thermodynamical quantities
rather than by the microscopic degrees of freedom, because the latter
can become infinitely large and, thus, meaningless for small
$\gamma$'s. For a discussion of (multi-)polytropical indices see
\citet{Livadiotis-2016}.  Nevertheless, since $\gamma$ in terms of $f$
is from a microphysical point of view more convenient than using a
thermodynamical representation, we continue to use $f=2/(\gamma-1)$
and call this ``equivalent degrees of freedom''.

As can be seen from Fig.~\ref{fig:1}, the spatial extent of the IHS along the
inflow axis $z$ decreases/increases with decreasing/increasing $\gamma$.
Thus, a $\gamma$ smaller than $5/3$ can explain the observed extent of
the IHS of roughly 30\,AU \citep{Izmodenov-etal-2014}.
However, simulations including neutrals and magnetic fields are required to
estimate the exact behavior. Nevertheless, as discussed by
\citet{Nicolaou-etal-2014}, \citet{Jacobs-Poedts-2011},
\citet{Kartalev-etal-2006}, and \citet{Farrugia-etal-2001}, the polytropic
index of the solar wind ranges from 1.4 to 2.1
although it should be noted that these authors neglect the behavior of PUIs;
others include temperature anisotropies or $\kappa$~distributions.
Including PUIs and following \citet{Fahr-Siewert-2015} or
\citet{Wu-etal-2009}, we easily obtain $\gamma > 5/3$, leading to an increase
of the IHAS's extent in a hydrodynamic single-fluid model.

For a single-fluid model with a parametric global $\gamma$, the compression
ratios change depending on the polytropic index. The standard \RH relation
defining $s$ gives
\begin{eqnarray}
  s  = \frac{(\gamma+1)  M^{2}_{1}}{(\gamma-1) M^{2}_{1} + 2}
\end{eqnarray}
\citep{Landau-Lifschitz-1972H}. In the limit of infinite Mach numbers, one
obtains the maximum compression ratio
\begin{eqnarray}
  \label{eq:s_max}
  s_{\textnormal{max}} := \lim_{M_1 \rightarrow \infty} s
  = \frac{\gamma + 1}{\gamma -1} \, .
\end{eqnarray}

From Fig.~\ref{fig:1}, it is also evident that the behaviors in the
upwind and tail directions are different: While in the upwind
direction the TS remains at the same position, in the tailward
direction the TS distance varies with the polytropic index.  For
$\gamma \ne 4/3$, the TS moves inward. In other words, of all cases
shown in Fig.~\ref{fig:1}, the TS in downwind direction is located
farthest from the Sun for $\gamma = 4/3$. This maximum is neither
theoretically nor empirically determined; it is just
the model with the most distant TS in the tail. The difference between
the tail and upwind directions is that in the latter, the supersonic
solar/stellar wind ram pressure balances that of the ISM, while in the
tail direction, it only has to balance the thermal pressure.

In the upwind direction, the TS position is independent of $\gamma$
because the characteristic along the inflow axis is the momentum
balance between the total momentum of the ISM and that of the supersonic
stellar/solar wind, which in the chosen parameter range is dominated
by the ram pressures of both media.  Note that the thermal pressure is
irrelevant. Thus, the TS distance is \citep{Parker-1963}
\begin{eqnarray}
  r_{\rm TS} = r_{0} \, 
  \frac{v_{\rm sw}}{v_{\rm ism}}\sqrt{\frac{\rho_{\rm sw}}{\rho_{\rm ism}}}
\end{eqnarray}
with $r_{0}=1$\,AU.

Concluding this section, we state that with different $\gamma$ values
the compression ratio and, furthermore, the thickness of the HAS can
be changed.  Although we have discussed only changes of the IHAS, our
results also apply to the OHAS analogously.  Deviations of the
polytropic index from the mono-atomic $\gamma=5/3$ can be caused, for
example, by a mixture of mono-atomic species (like protons) with more
complex atoms or by including magnetic fields.

Furthermore, the solar wind polytropic index is expected to change
with the solar cycle according to \citet{Nicolaou-etal-2014} and
\citet{Jacobs-Poedts-2011}. This will affect the shock structures, but
to our knowledge, such a time-dependent model does not yet exist.
Nonetheless, a problem with simulating the helio-/astrospheres with
globally constant polytropic indices is that the IHAS can shrink but
at the cost of an increasing compression ratio, which is not observed
for the heliosphere.  The idea to avoid this problem is to consider
changes of the polytropic index after a shock passage of the solar
wind. We discuss this in the following and see that the problem of
compression ratios being too large remains.
~\\

\section{Variable polytropic indices across a shock}
\label{sec:4}

In this section, we discuss the \RH relations for the density, pressure, 
and temperature ratios for systems containing shocks with different upstream
and downstream polytropic indices. To this end, we consider the general case
of polytropic indices $\gamma_i \in \, ]1,3]$. Then, for the application
to the helio-/astrospheres, we fix the upstream polytropic
index to $\gamma_{1}=5/3$. 

In the following, only the main results are stated, while technical
details are given in the Appendix.  Various ratios have already been
discussed by \citet{Drake-2006} and similarly by
\citet{Nieuwenhuijzen-etal-1993}.  The explicit form of the
compression ratio given by the latter authors leads, however, in the
hypersonic limit  of high upstream Mach numbers ($M_{1}\gg1$) to
unexpected results. In this limit, one expects a compression ratio of
the form (\ref{eq:s_max}) with $\gamma=\gamma_{2}$ (see
Eq.~(\ref{eq:a11}) in the Appendix).  This is already discussed in
\citet{Drake-2006}, where, however, the various solution branches were not
considered. Therefore, we newly perform these calculations following the
notation in textbooks like \citet{Courant-Friedrichs-1948},
\citet{Landau-Lifschitz-1972H}, or \citet{Goedbloed-etal-2010}.
\begin{figure}[t!]
\includegraphics[width=0.995\columnwidth]{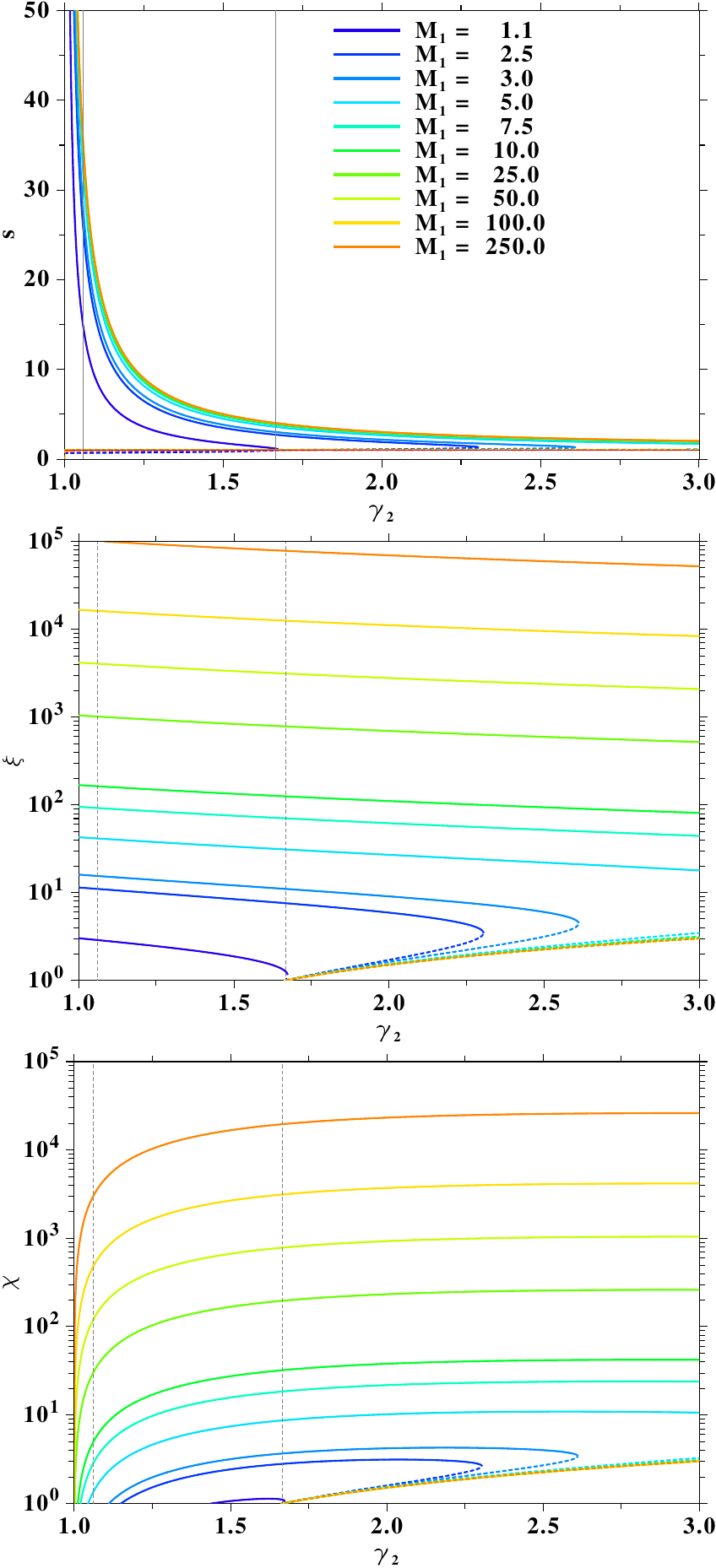}
  \caption{ \label{fig:2} The ratios $s^{\pm}$ (upper panel, in linear
    scale), $\xi^{\pm}$ (middle panel, in logarithmic scaling), and
    $\chi^{\pm}$ (bottom panel, in logarithmic scaling) for a fixed
    $\gamma_{1}=5/3$ and selected values of $M_{1}$ (as indicated in
    the legend of the upper panel) as a function of $\gamma_2$, with 1
    for isothermal conditions (infinite degrees of freedom) and 3 for
    one degree of freedom. The two vertical lines mark
    $\gamma_{2}= 1.06$ \citep{Izmodenov-etal-2014}, and
    $\gamma_{1}=\gamma_{2}=5/3$.  The solid lines correspond to the
    positive solutions $s^{+},\xi^{+}, \chi^{+}$, the dotted lines to
    the negative ones $s^{-},\xi^{-}, \chi^{-}$.  }
\end{figure}

The conservation of energy across a shock is given by
\begin{equation}
  \label{eq:1}
  w_{1}-w_{2} + \frac{1}{2}(V_{1}+V_{2})(P_{2}-P_{1}) = 0
\end{equation}
with the enthalpies
\begin{eqnarray}
  \label{eq:2}
  w_{i} = \frac{\gamma_i P_{i} V_{i}}{\gamma_i - 1}
\end{eqnarray}
and specific volumes $V_i := 1/\rho_i$.
To include oblique shocks, we use $v_{{\rm n},i}$ to denote the
velocity components along the shock normal vector, and the normal Mach number
$M_{\rm n,1} := v_{\rm n,1}/c_{\rm s,1} = M_{1} \sin\vartheta$ with the shock
angle $\vartheta$, i.e., the angle between the shock and the inflow velocity.
After a short calculation, we find for the compression ratio
$s=\rho_{2}/\rho_{1}=v_{\rm n,1}/v_{\rm n,2}$ in the rest frame of the shock 
\begin{equation}
  \label{eq:3}
  s = s^+ = \frac{M_{\rm n,1}^2 \gamma_1
    (\gamma_2 + 1)}{(M_{\rm n,1}^2 \gamma_1 + 1) \gamma_2 - \Gamma_{1 2}} \ ,
\end{equation}
and for the pressure ratio $\xi= P_{2}/P_{1}$
\begin{equation}
  \label{eq:4}
  \xi = \xi^+ = 
  \frac{1}{\gd+1} \, \left( M_{\rm n,1}^{2}\gu + 1 + \Gamma_{12} \right) \ ,
\end{equation}
where
\begin{eqnarray}
  \label{eq:3a}
  \Gamma_{12} := \sqrt{\gu^{2} M_{\rm n,1}^{4}
    + \dfrac{2\gu M_{\rm n,1}^{2}(\gu-\gd^{2})}{\gu-1}+\gd^{2}} \ .
\end{eqnarray}
The superscript ``$+$'' refers to the positive solution branch, as
discussed in the Appendix.

Assuming that the temperature obeys the ideal gas law
$P_i \propto \rho_i T_i$, we obtain for the temperature ratio $\chi=T_{2}/T_{1}$
\begin{eqnarray}
  \label{eq:5}
  \chi = \chi^+ =
  \frac{P_{2} \, \rho_{1}}{P_{1} \, \rho_{2}} = \frac{\xi^{+}}{s^{+}} \ .
\end{eqnarray}
The respective negative branches $s^{-}$, $\xi^{-}$, and $\chi^{-}$ are
disregarded for physical reasons; this is discussed in detail in 
Subsection~\ref{negsol}.

In the hypersonic limit ($M_{\rm n,1}\gg1$), the ratios (\ref{eq:3}),
(\ref{eq:4}), and (\ref{eq:5}) become
\begin{eqnarray}
  \label{eq:7}   s  &\simeq& \frac{\gd+1}{\gd-1} \\
  \label{eq:8} \xi  &\simeq& \frac{2 \gu}{\gd+1} \, M_{\rm n,1}^{2} \\
  \label{eq:9} \chi &\simeq& \frac{2  \gu (\gd-1)}{(\gd+1)^{2}} \, M_{\rm n,1}^{2}\, .
\end{eqnarray}
From Table~\ref{tab:1} and the shapes of the functions (\ref{eq:3}),
(\ref{eq:4}), and (\ref{eq:5}), one can see that for our simulations the
hypersonic approximation is reasonable at the TS.  Note that the radicand of
$\Gamma_{12}$ in Eq.~(\ref{eq:3a}) can become negative for
\begin{equation}
  \label{eq:9b}
  \gamma_2 < \gamma_{1} \, M_{\rm n,1} \sqrt{ \frac{(\gamma_{1}-1)
    M_{\rm n,1}^{2}+2}{2\gamma_{1} M_{\rm n,1}^{2}-\gamma_{1}+1} } \,
 \, ,
\end{equation}
which leads to complex values for the ratios and, thus, non-physical
solutions (see Table~\ref{tab:2}).  For $\gu=\gd$, the above ratios
reproduce the well-known results.

\begin{table}[t]
  \centering
  \begin{tabular}{ccc|cccc} 
    $\gamma_{1}$ & $\gamma_{2}$ & $M_{1}$ & $s$ & $\xi$ & $\chi$ & $\zeta$ \\
    \hline
    5/3 & 1.06 & 1.001 & 13.5 &  2.55 &  0.19 & 10.7 \\
    5/3 & 1.06 & 5.000 & 31.3 & 41.34 &  1.32 & 71,0 \\
    5/3 & 4/3  & 1.001 &  2.5 &  2.00 &  0.89 & 5.9 \\
    5/3 & 4/3  & 5.000 &  6.3 & 36.08 &   5.7 & 30.8 \\ 
    4/3 & 5/3  & 1.001 &  \multicolumn{4}{c}{--- all ratios complex ---} \\
    4/3 & 5/3  & 5.000 & 3.1  & 23.65 &  7.58 & 0.36 
  \end{tabular}
  \caption{\label{tab:2} \rm \begin{flushleft}
      The respective ratios of density $s=s(\gamma_{1},\gamma_{2},M_{1})$,
      pressure $\xi=\xi(\gamma_{1},\gamma_{2},M_{1})$, temperature
      $\chi=\chi(\gamma_{1},\gamma_{2},M_{1})$, and the ratio of the
      entropy parameter $\zeta=\zeta(\gamma_{1},\gamma_{2},M_{1})$ (see Eq.~(\ref{eq:enpa}) below) 
       for specific values of
      upstream and downstream polytropic indices $(\gamma_{1}, \gamma_{2})$
      and Mach number $M_{1}$. Allowed solutions are marked by ratios
      $s$, $\xi$, $\chi$, and $\zeta$ being real and greater than unity.
    \end{flushleft}}
\end{table}

In Fig.~\ref{fig:2}, both solutions for the ratios
$s^{\pm}, \xi^{\pm}$, and $\chi^{\pm}$ are shown.  The compression
ratio $s^{+}$ increases with decreasing $\gamma_{2}$, which becomes
particularly evident in the hypersonic limit (\ref{eq:7}). The
negative solution $s^{-}$ is slightly larger than one for
$\gamma_{2}>\gamma_{1}$ and slightly lower one for
$\gamma_{2}<\gamma_{1}$. The pressure ratio $\xi^{+}$ grows rapidly
with increasing Mach number. For low Mach numbers, it decreases with
increasing $\gamma_{2}$ until $\Gamma_{12}$ becomes imaginary.  An
increase beyond the value determined by Eq. (\ref{eq:9b}) (e.g.,\
$\gamma_{2}<2.02$ for $\gamma_{1}=5/3$ and $M_{1,n}=2$) is not
possible. For sufficiently high Mach numbers (i.e., $M_{\rm n,1}>3.6$
for $\gamma_{1}=5/3$ and $\gamma_{2}=3$) all solutions are real in the
domain of interest. For constant Mach numbers and given $\gamma_{1}$,
the pressure ratio decreases with increasing $\gamma_{2}$.
Considering the ratios $s^+,\xi^+$, and $\chi^+$, almost all
$\gamma_1 \rightarrow \gamma_2$ transitions for high Mach numbers are
allowed.

\begin{figure}[t!]
\includegraphics[width=0.995\columnwidth]{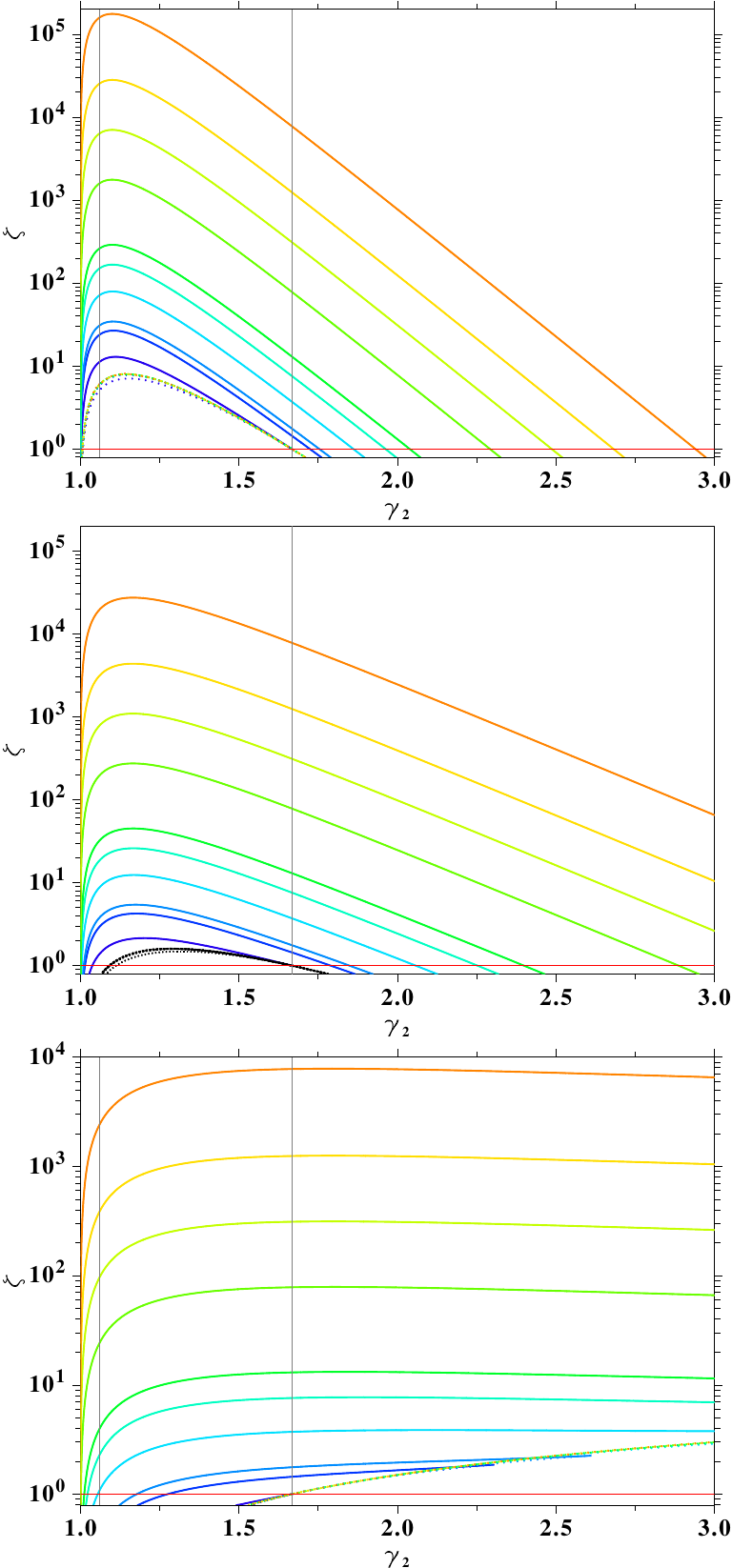}
  \caption{The entropy ratio $\zeta$ for the positive solutions
    $\xi^{+}$ and $s^{+}$ (solid lines) and the negative solutions
    $\xi^{-}$ and $s^{-}$ (dotted lines).  The upper panel shows the
    solution for $\tau=T_{1}/T_{0}=10^{2}$, the middle panel for
    $\tau=10$, and the lower panel for $\tau=1$.  The red
    horizontal lines indicate the cutoff value $\zeta=1$ (cf.\
    Eq.~(\ref{eq:as7})). The color coding is identical to that of
    Fig.~\ref{fig:2}. \label{fig:3}}
\end{figure}


Note that also the negative solutions $s^-,\xi^-$, and $\chi^-$ can
assume real values. Therfore, for given $\gamma_{1}$ and $\gamma_{2}$
a criterion for the selection of the correct solution is required. As
shown in the following section, this criterion is provided by the
entropy jump condition and the maximum entropy production principle
 that, to the best of our knowledge, has not been discussed in the
literature in this context.
~\\

\section{The multi-polytropic entropy difference
  across a shock}
\label{sec:5}

\subsection{The absolute entropy for arbitrary degrees
  of freedom}

To determine the entropy difference across a shock, it seems plausible
to use the fundamental thermodynamic relation 
\begin{eqnarray}
  \label{eq:ts}
  \dd \tilde{S} = \frac{k}{2 T} \dd (f T) + \frac{p}{T} \dd V 
\end{eqnarray}
with volume $V$ (not to be confused with the specific volume
$V_i= 1/\rho_i$) and a variable degree of freedom $f$.  Direct
integration of Eq.~(\ref{eq:ts}) leads to a non-local expression for
the entropy, namely a line-integrated entropy across the shock. Since
we are interested in an expression for the entropy that can be
evaluated locally, we make use of the absolute entropy $\tilde{S}$
derived by \citet{Syrkin-1924} for a gas with arbitrary degrees of
freedom given by
\begin{eqnarray}
  \label{eq:as1}
  \tilde{S} = k N \ln\left(\frac{(2\pi m_0 k)^{f/2} T^{f/2} a_0^{(f-3)}V
      e^{(1+f/2)}}{N h^{f}}\right) \, ,
\end{eqnarray}
where $N$ is the number of particles, $m_0$ and $a_0$ are the particles' mass
and diameter, $h$ is the Planck constant, and $e = \exp(1)$. Note that the
argument of the logarithm is dimensionless.

For later use, we rewrite Eq.~(\ref{eq:as1}) in a form with various
single logarithms which may carry dimensions. This is guided by the
exponents occurring in Eq.~(\ref{eq:as1}).  Further, we define the
dimensionless entropy per particle $S := \tilde{S}/(k N)$, which can
be expressed as
\begin{eqnarray}
  \label{eq:as2}
  S &=& \frac{f}{2} \left[
    \ln T + \ln \left( \frac{2\pi e k m_0 a_0^2 }{h^{2}}  \right) \right]
  \\\nonumber &&  - \left[ \ln (n) + \ln \left(\frac{a_0^3}{e} \right) \right] 
\end{eqnarray}
using $n=N/V$. The constant
\begin{eqnarray}
  \label{eq:as3}
  \frac{2\pi e k \, m_0 a_0^2}{h^{2}} =: \frac{1}{T_{0}} \ ,
\end{eqnarray}
i.e., the argument of the second logarithm, has the dimension of an
inverse temperature $T_{0}$, and, thus, cancels the dimension of the
argument in the first term. Likewise,
\begin{eqnarray}
  \label{eq:def_n0}
  \frac{a_0^3}{e} =: \frac{1}{n_{0}}
\end{eqnarray}
has dimension of a volume (or inverse number density), and cancels the
dimension of $\ln(n)$. The absolute entropy (\ref{eq:as2}) now assumes
the rather compact form
\begin{eqnarray}
  \label{eq:as5}
  S &=& \frac{f}{2} \ln \left( \frac{T}{T_0} \right)
  - \ln \left( \frac{n}{n_0} \right) \ .
\end{eqnarray}

In the next step, the constants $T_0$ and $n_0$, which depend on the
particle mass $m_0$ and diameter $a_0$ are fixed. While the proton
mass $m_{\rm p}$ is easily identified as a suitable choice for $m_0$,
it would not be appropriate in this context to use the proton
``diameter'' $a_{\rm p}$ for $a_0$ for the following reason: In Syrkin's
derivation of formula (\ref{eq:as1}), the particle diameter appears as
a measure of the minimum distance between adjacent particles. But in a
plasma, it is from a physical point of view not possible to place two
protons directly side by side.  Choosing the hydrogen atom diameter
is, thus, more appropriate because a plasma is always quasi-neutral.
Since a plasma at low temperatures tends to recombine and neutralize,
we, however, consider a somewhat larger diameter given by the
classical Bohr radius, which enables the plasma to stay in its
quasi-neutral state, i.e., a mixture of protons and electrons (and not
a (hydrogen) gas).  Therefore, with the Bohr radius of the hydrogen atom
$r_{\rm B}=5.291\times10^{-9}$\,cm, we obtain after normalizing
Eqs.~(\ref{eq:as3}) and (\ref{eq:def_n0}) to $m_0=m_{\rm p}$ and
$a_0=a_{\rm B}= 2 r_{\rm B}$,
\begin{eqnarray}
  \label{eq:as32}
  T_0 &=& (99.5 \ {\rm K})
  \left(\frac{m_0}{m_{\rm p}}\right)^{-2}
  \left(\frac{a_0}{a_{\rm B}}\right)^{-1} \\
  n_0 &=& (2.3 \times 10^{24} \ {\rm cm}^{-3})
  \left(\frac{a_0}{a_{\rm B}}\right)^{-3} \ .
\end{eqnarray}
In the following, we approximate $T_{0} \approx 100$~K.

As an example, we compute the total entropy of a dense H$_{2}$
molecular cloud with $T=100$~K and $n=100$~cm$^{-3}$ assuming that at
this temperature all degrees of freedom except those of translational
motion are frozen-in (implying $f=3$). We approximate the radius and
mass of the H$_2$ molecule by twice the values for a single hydrogen
atom. We obtain an absolute normalized entropy per particle of
$S \approx 53$.  Note that the absolute entropy per particle
(\ref{eq:as5}) is dominated by the normalization density $n_{0}$ for a
wide range of astrophysical applications, for example, ISM or
astrospherical shocks, because the temperature is usually below
$T<10^{9}$\,K. For other scenarios, like stellar cores, the density $n$
can easily become larger than $n_{0}$.
~\\

\subsection{Entropy difference for arbitrary degrees
  of freedom}

We now consider variable degrees of freedom through a shock 
passage and calculate the entropy difference as the difference of the
absolute entropies (\ref{eq:as5}) from each side. 

The conservation law for the specific entropy with the fluid velocity
$\vec{v}$ reads
\begin{eqnarray}
  \label{eq:e1}
  \frac{\partial (\rho S)}{\partial t} + \nabla \cdot (\rho \vec{v} S) = 0
  \, .
\end{eqnarray}
Because a shock is an irreversible transition, the
corresponding \RH condition \citep{Goedbloed-etal-2010} is
\begin{eqnarray}
  \label{eq:e2}
  -u [S_{1}-S_{2}] + \rho_{1} v_{\rm n,1} S_{1} - \rho_{2} v_{\rm n,2} S_{2} \le 0 \, ,
\end{eqnarray}
where $u$ is the shock speed and $v_{\rm n}$ its component normal to the
shock. Using the equation of continuity
$\rho_{1} v_{\rm n,1}=\rho_{2}v_{\rm n,2}$, this leads for a
stationary shock ($u=0$) to
\begin{eqnarray}
  \label{eq:e3}
  S_{2} - S_{1} \ge 0 \, .
\end{eqnarray}
Allowing for different degrees of freedom before and after the shock, we can
compute the normalized entropy difference up- and downstream of a shock
by means of Eq. (\ref{eq:as5}). We find
\begin{eqnarray}
  \label{eq:as62}
  S_2 - S_1 &=& \frac{f_{2}}{2} \ln\left(\frac{T_{2}}{T_{1}}\right)  -  
  \ln\left(\frac{n_{2}}{n_{1}}\right)  \\ \nonumber
  && + \frac{f_{2}-f_{1}}{2}  \ln\left(
    \frac{T_{1}}{T_{0}}\right) \ge 0 \, .
\end{eqnarray}

Keeping in mind that for some applications like astrospherical shocks
we may also encounter different (e.g., dissociative) species on both
sides of the shock, we need to use different values for $(m_0,a_0)$,
and, therefore, different values for $(T_0, n_0)$. In this case, the
following term has to be added to the right-hand side of
Eq.~(\ref{eq:as62})
\begin{eqnarray}
  \label{eq:add_T01} \nonumber
  && - \frac{f_2}{2} \ln \left( \frac{T_{0,2}}{T_{0,1}} \right)
  + \ln \left( \frac{n_{0,2}}{n_{0,1}} \right) \\
  &=& \frac{f_2}{2} \ln \left( \frac{m_{0,2}}{m_{0,1}} \right)
  + \left(f_2 - 3 \right) \,
  \ln \left( \frac{a_{0,2}}{a_{0,1}} \right) \, .
\end{eqnarray}
Then, however, we cannot use only the difference of the normalized
entropies (\ref{eq:as62}), but have also to take into account
different numbers of particles $N_{1} \ne N_{2}$ and additional mixing
terms. For our application, we assume that we have the same kind of
particles on both sides of the shock.

Requiring the validity of the ideal gas law on both sides of the
shock, formula (\ref{eq:as62}) can easily be reformulated in terms of
the ratios $\chi$ and $s$ as
\begin{eqnarray}
  \label{eq:as63}
  S_2 - S_1 &=& \frac{f_{2}}{2} \ln (\chi) - \ln (s) + \frac{f_{2}-f_{1}}{2}
  \ln \left( \frac{T_{1}}{T_{0}}\right) \ge 0 \, . \quad\quad
\end{eqnarray}
We remark that both ratios are functions of the upstream parameters
$M_{1}$ and $f_{1}$ as well as the downstream degree of freedom $f_{2}$,
which is assumed to be independent of the upstream values. 
However, for the excitations of higher degrees of freedom, e.g., by
rotations or vibrations, the downstream temperature needs to be high 
and, thus, $f_{2}$ will depend on the underlying physical process.

In terms of the pressure ratio $\xi$ and the polytropic indices
$\gamma_{1},\gamma_{2}$, the inequality (\ref{eq:as63}) reads
\begin{eqnarray}
  \label{eq:as64}
  \frac{\ln(\xi) - \gamma_{2} \ln (s)}{\gamma_{2}-1}
   +\frac{\gamma_{1}-\gamma_{2}}{(\gamma_{1}-1)(\gamma_{2}-1)}
  \ln \left( \frac{T_{1}}{T_{0}} \right) \ge 0
\end{eqnarray}
using the definition of the equivalent degrees of freedom $f_i = 2/(\gamma_i-1)$.
In order to obtain the ratio 
\begin{eqnarray}
  \label{eq:enpa}
  \zeta := \left(\exp(S_2)/\exp(S_1)\right)^{\gamma_{2}-1} \ , 
\end{eqnarray}
we take the exponential of~(\ref{eq:as64}) and get
\begin{eqnarray}
\label{eq:as7}
  \zeta = \xi \, s^{-\gamma_{2}}
  \left( \frac{T_{1}}{T_{0}}
                    \right)^{\frac{\gamma_{1}-\gamma_{2}}{\gamma_{1}-1}}
                    \ge 1 \, .
\end{eqnarray}
In terms of the degrees of freedom and the temperature ratio, this relation 
yields
\begin{eqnarray}
  \label{eq:as8}
\chi^{f_2} s^{-2} 
  \left( \frac{T_{1}}{T_{0}} \right)^{f_2-f_1} \ge 1 \, .
\end{eqnarray}
Note that the right hand side differs from unity only if degrees of
freedom change across the shock. 

In Fig.~\ref{fig:3}, the ratio $\zeta$ as function of $\gamma_{2}$ for
a fixed $\gamma_{1}$ is shown for three different values of the ratio
$\tau := T_{1}/T_{0}$, namely $\tau = 100$, $\tau=10$, and $\tau = 1$.
For $\tau=10,100$, the entropy ratio first increases rapidly for all
displayed Mach numbers, and after reaching a maximum, it decreases to
values below unity for $M<250$.  The increase of the entropy ratio for
$\tau=1$ is also quite fast for small values of $\gamma_{2}$.  For
larger values of $\gamma_{2}$ the entropy ratio becomes approxmately
constant until it finally slowly decreases. This is the limiting case
for physical solutions because then the plasma freezes out and we
obtain recombined neutral atoms. We point out that in (\ref{eq:as8})
the case $\tau=1$ is not identical to the case where
$\gamma_{1}=\gamma_{2}$ because $s$ and $\chi$ depend both on
$\gamma_{1}$ and $\gamma_{2}$. Moreover, for all $tau$'s the entropy
ratios resulting from the negative roots $\xi^-$ and $s^-$ are also
allowed because they can be larger than unity.

\subsection{Discussion of the negative roots}
\label{negsol}
The solutions for the compression and pressure ratios (and, thus, the
temperature and entropy ratios) consist each of two branches
$s^{\pm}$, $\xi^{\pm}$, which depend on the sign in front of 
$\Gamma_{12}$ (see Eq.(\ref{SOLxi})). 

The positive solutions represent the standard compression shocks,
whereas the negative solutions are rarefaction shocks as discussed in
the literature for special ``fluids'', i.e., so-called BZT fluids
(Bethe-Zeldovitch-Thompson); for details see \citet{Cramer-Park-1999}.
Here, the rarefaction shocks must be excluded because of the maximum
entropy production principle \citep{Martyushev-2013}: In all cases
where both solutions are allowed simultaneously, the entropy
production for the positive solutions is always larger as for
the negative roots (cf. Fig.~4). 
~\\

\subsection{Application to the helio-/astrospheres}

In our studies of the helio- and astrospheres, variable polytropic
indices have been used.  We have calculated for specific values of
$\gamma_{1}$, $\gamma_{2}$, and $M_{1}$ the compression, pressure, and
temperature ratios as well as the entropy difference (see
Table~\ref{tab:2}). Some ratios become complex-valued, which is caused
by a negative radicand of the square root~(\ref{eq:3a}) and, thus, not
all values of $\gu,\gd$, in addition to the cases where $\zeta<1$,
lead to physical solutions. Therefore, variable polytropic indices
across a (termination) shock can easily lead to unrealistic physical
conditions.  This is particularly true for large-scale
helio-/astrospherical simulations, where in the flank regions the Mach
numbers can become small.
~\\

\section{Discussion and conclusions}
\label{sec:6}

In this work, we have derived multi-polytropic generalized \RH relations.
We gave explicit expressions for the ratios of the densities, pressures, and
temperatures across a shock. The derivation included, for the first time,
a consistent consideration of the associated change in entropy.

For the examples of the helio-/astrospheres, we have discussed, on the
one hand, the case of different globally constant polytropic indices
and, on the other hand, a changing polytropic index in a shock
transition. The polytropic index in the solar wind is varying and
usually different from $5/3$.  For smaller values of $\gamma$ in the
entire integration region (also in the ISM), the thickness of the IHAS
and OHAS in upwind direction shrinks, while for higher values, it
grows. In the former case, the compression ratio increases, while in
the latter it decreases. A decrease in the compression ratio is needed
in order to explain the {\em Voyager} observations, while an increase
for small $\gamma$'s is incompatible with the observations.

Since the compression ratio, which is an important quantity for the
acceleration of particles, is a function of the polytropic index, it
is of great interest to study astrospheres under varying polytropic
indices in order to analyze the acceleration of energetic particles.

The interaction of astrospheres, whose abundances are not
solar-like, with the ISM can lead to a change in the polytropic index
beyond the (bow) shock passage. This is particularly true when a star
is born in a cold H$_{2}$ cloud, where the diatomic hydrogen
dissociates after the shock passage. Relation~(\ref{eq:as63}) can be
adapted to this case, taking into account the different numbers of
particles $N_{1}$, $N_{2}$, and possible mixing terms, however, it
will require a multi-fluid description of the shock transition region.

Thus, a detailed study of multi-fluid models with different polytropic
indices not only across the shock, but for example also for different
species like for solar/stellar wind protons and PUIs is needed. The
problem at the moment is that to our knowledge the available
simulation codes cannot yet handle such a setting. Further, it is not
clear if such a configuration is at all thermodynamically possible for
fluids (see \citet{McKee-Holliman-1999} for multi-pressure polytropes
to model the structure and stability of molecular clouds).

We point out that the above study is not restricted to the applications 
	 of helio- and astrospheres but concerns all scenarios in which the polytropic 
	 index varies over a hydrodynamic shock, e.g., interstellar shocks
  \citep[for example][]{Pittard-Parkin-2016} or supernova explosions
  \citep[for example][]{Bolte-etal-2015}.

  For MHD shocks, the situation is more complicated because the
  jump conditions are characterized by a three-parameter family (the normal
  alfv\'enic Mach number, the ratio between the thermal and magnetic
  field pressure, and the angle between the inflow and magnetic field
  vector), whereas for hydrodynamical shocks, they are characterized by
  a one-parameter family (the normal Mach number). The more
  challenging task of MHD shocks with varying polytropic indices will be
  addressed in a future study.

~\\

\begin{acknowledgements} 
  KS, HF, and JK are grateful to the
  \emph{Deut\-sche For\-schungs\-ge\-mein\-schaft (DFG)} funding the
  projects SCHE334/9-1, SCHE334/9-2, and FI706/15-1. \\
\end{acknowledgements}

\bibliographystyle{apj}

\appendix
\section{Generalized mulit-polytropic Rankine-Hugoniot
  relations}\label{app:1}
\subsection{The pressure ratio $\xi$}

Substituting the enthalpy (\ref{eq:2}) into the equation of energy
conservation (\ref{eq:1}) gives
\begin{eqnarray}
  \frac{\gu P_{1} V_{1}}{\gu - 1} - \frac{\gd P_{2} V_{2}}{\gd - 1}
  + \frac{1}{2}(V_{1}+V_{2})(P_{2}-P_{1}) = 0 \, .
\end{eqnarray}
From this, it immediately follows that
\begin{eqnarray}
  \label{eq:a1a}
  \frac{V_{1}}{V_{2}} =
  \frac{ (\gd-1)P_{1} + (\gd+1)P_{2}}
  { (\gu+1)P_{1} + (\gu-1)P_{2}}  \times \frac{\gu-1}{\gd-1} \, .
\end{eqnarray}
Furthermore, using Eq.~(\ref{eq:a1a}) and the pressure ratio
$\xi=P_{2}/P_{1}$, the square of the mass current $j$ becomes
\begin{eqnarray}
  \label{eq:a2}
  j^{2} &=& \frac{P_{2}-P_{1}}{V_{1}-V_{2}} =
            \dfrac{P_{1}}{2 V_{1}}(\xi-1) \, 
            \frac{\left[\gd-1 + (\gd+1)\xi\right]\left[\gu-1\right]}
            {- \gd + 1 + (\gu-1)\xi} \, .
\end{eqnarray}
 With the velocities $v_{i}=j V_{i}$ and the Mach numbers
$M_{i}=v_{i}/c_{\rm s,i}$, we obtain from Eq.~(\ref{eq:a2}) and the 
defining relation for the upstream sound speed
\begin{equation}
  c_{\rm s,i} = \sqrt{\gamma_i P_i V_i}
\end{equation}
after some algebraic manipulations
\begin{eqnarray}\label{SOLxi}
  \nonumber
  \frac{2 M_{1}^{2} \, \gu}{\gu - 1} &=& (1-\xi) \, 
  \frac{\gd-1 + (\gd+1)\xi}
  {\gd-1- (\gu-1)\xi}
  \\\label{eq:a5} \Leftrightarrow \, \xi^{\pm} &=& \frac{1}{\gd + 1} \, \bigl( 
  M_{1}^{2}\gu + 1 \pm  \Gamma_{1 2} \bigr) \, ,
\end{eqnarray}
where the superscripts  $\pm$ denote the two roots and 
\begin{equation}
  \Gamma_{1 2} := \sqrt{\gu^{2}M_{1}^{4}+2\gu M_{1}^{2} \,
    \frac{\gu-\gd^{2}}{\gu-1}+\gd^{2}} \, .
\end{equation}
Setting $\gu = \gd = \gamma$ yields
\begin{equation}
  \label{eq:a6}
  \left( \xi^{+},  \xi^{-} \right) = \left( \frac{ 2\gamma M_{1}^{2} - \gamma + 1}{\gamma + 1} \,
    , 1 \right) \, ,
\end{equation}
which is the usual \RH relation for $\xi^{+}$ and that for rarefaction
shocks for $\xi^{-}$ (see Section~5.3).
~\\

\subsection{The compression ratio}

Rewriting Eq.~(\ref{eq:a2}), we obtain for the square of the mass current
\begin{equation}\label{eq:a55}
  j^{2} = \frac{P_{1} (\xi - 1)s}{V_{1}(s - 1)} \, ,
\end{equation}
where $s = V_1/V_2$ is  the compression ratio. On the other hand,
using the expressions for $v_{1}$, $M_{1}$, and $c_{\rm s ,1}$, we
find
\begin{equation}
  \label{eq:a56}
  j^{2} = \frac{M_{1}^{2} \gamma_{1} P_{1}}{V_{1}} \, .
\end{equation}
Combining Eqs.~(\ref{eq:a55}) and~(\ref{eq:a56}) leads after some
algebra to
\begin{equation}
\label{eq:a61}
  s = \frac{M_1^2 \gamma_1}{M_1^2 \gamma_1 + 1 - \xi} \, .
\end{equation}
Substituting the solutions (\ref{eq:a5}) for the pressure ratio, we find
\begin{eqnarray}\label{eq:a91}
  s^{\pm} = \frac{M_1^2 \gamma_1 (\gamma_2 + 1)}{(M_1^2 \gamma_1 + 1) \gamma_2
    \mp \Gamma_{1 2}} \, .
\end{eqnarray}
In the case $\gamma_{1}=\gamma_{2}=\gamma$, these simplify to
\begin{equation}
  \left( s^{+}, s^{-} \right)  = \left(
    \frac{(\gamma + 1) M_{1}^{2}}{(\gamma-1) M_{1}^{2}+2}, 1 \right) \, ,
\end{equation}
where the solution $s^{+}$ is again the standard \RH compression ratio.
~\\

\subsection{The temperature ratio $\chi$}

The temperature ratio $\chi$ is given by
\begin{eqnarray}
  \label{eq:a12}
  \chi &=& \frac{T_{2}}{T_{1}} = \frac{P_{2}}{P_{1}}
  \frac{\rho_{1}}{\rho_{2}} = \frac{\xi}{s} \ ,
\end{eqnarray}
which can be easily calculated in terms of Eqs.~(\ref{SOLxi}) and
(\ref{eq:a91}).
~\\

\subsection{The shocked Mach number}

From Eq.~(\ref{eq:1}), the shocked Mach number $M_{2}$ becomes
\begin{eqnarray}
  \label{eq:a13}
  M_{2}^{2} = \frac{1}{\gamma_{2}}\frac{\gamma_{1} M_{1}^{2}}{s +
  \gamma_{1}M_{1}^{2}(s-1)} \, .
\end{eqnarray}

\subsection{Hypersonic limits}

The hypersonic limits of the pressure, compression, and temperature
ratios as well as that of the shocked Mach number are computed by
considering large upstream Mach numbers $M_1 \gg 1$ in
Eqs.~(\ref{SOLxi}), (\ref{eq:a91}), and (\ref{eq:a12}) for the
relevant values of $\gamma_1, \gamma_2$, resulting in the expressions
\begin{eqnarray}
  \label{eq:a11}
  \left( \xi^+, \xi^- \right) &\simeq& \left(
    \frac{2\gu}{\gd+1} M_{1}^{2} \, , \
    \frac{\gamma_2 - 1}{\gamma_1 - 1} \right) \\ \nonumber\\
  \left( s^+, s^- \right) &\simeq&  \left(
    \frac{\gd+1}{\gd-1} \, , 1 \right) \\ \nonumber \\
  \left( \chi^+, \chi^- \right) &\simeq& \left(
    \frac{2 \gu (\gd-1)}{(\gd+1)^{2}} M_{1}^{2} \, , \
    \frac{\gamma_2 - 1}{\gamma_1 - 1} \right) \\ \nonumber\\
  \left(M_{2}^{+},M_{2}^{-} \right)  &\simeq&  \left( 
 \sqrt{\frac{\gamma_{2}-1}{2\gamma_{2}}} \, , \ 
 \sqrt{\frac{\gamma_{1}}{\gamma_{2}}}M_{1} \right) \, .
\end{eqnarray}
Note that the compression ratio $s^{+}$ is independent of $\gu$. Moreover,
in the case $\gu=\gd$, the standard \RH relations for high Mach
numbers (positive roots) are recovered.

\subsection{Oblique shocks}

Replacing the Mach number $M_{1}$ by $M_{1} \sin\vartheta$ and $v_i$ by
$v_{{\rm n},i} = \vec{n}\cdot\vec{v}_i$, also oblique shocks can be studied.
Here, $\vec{n}$ is the shock normal vector and $\vartheta$
the angle between the shock and the inflow velocity.

\end{document}